\documentclass[sigconf]{acmart}

\usepackage{listings}

\usepackage{xspace}

\usepackage{multirow}

\usepackage{commands}

\usepackage{subcaption}

\AddToHook{env/lstlisting/begin}{\fboxsep=0pt}

\usepackage{code-styles}

\AtBeginDocument{%
  }

\copyrightyear{2026}
\acmYear{2026}
\setcopyright{cc}
\setcctype{by}
\acmConference[ICSE-NIER '26]{2026 IEEE/ACM 48th International Conference on Software Engineering}{April 12--18, 2026}{Rio de Janeiro, Brazil}
\acmBooktitle{2026 IEEE/ACM 48th International Conference on Software Engineering (ICSE-NIER '26), April 12--18, 2026, Rio de Janeiro, Brazil}
\acmPrice{}
\acmDOI{10.1145/3786582.3786799}
\acmISBN{979-8-4007-2425-1/2026/04}

\begin{document}
\title{Build Code is Still Code: \\ Finding the Antidote for Pipeline Poisoning}

\newcommand{\ucf}{\affiliation{%
		\institution{University of Central Florida}
		\city{Orlando}
		\state{Florida}
		\country{USA}
	}
}

\author{Brent Pappas}
\email{brent.pappas@ucf.edu}
\orcid{0009-0003-0780-743X}
\ucf{}

\author{Paul Gazzillo}
\email{paul.gazzillo@ucf.edu}
\orcid{0000-0003-1425-8873}
\ucf{}

\renewcommand{\shortauthors}{Pappas and Gazzillo}

\begin{abstract}
    Open source C code underpins society's computing infrastructure.
Decades of work has helped harden C code against attackers, but C projects do
not consist of only C code.
C projects also contain build system code for automating development tasks like
compilation, testing, and packaging.
These build systems are critcal to software supply chain security and
vulnerable to being poisoned, with the XZ Utils and SolarWinds attacks being
recent examples.
%
%
Existing techniques try to harden software supply chains by verifying software
dependencies, but such methods ignore the build system itself.
Similarly, classic software security checkers only analyze and monitor program
code, not build system code.
%
Moreover, poisoned build systems can easily circumvent tools for detecting
program code vulnerabilities by disabling such checks.
%
%
We present development phase isolation, a novel strategy for hardening build systems
against poisoning by modeling the information and behavior permissions of build
automation as if it were program code.
%
%
We have prototyped this approach as a tool called \tool{}, which successfully
detects and warns about the poisoned test files involved in the XZ Utils
attack.
%
We outline our future plans to protect against pipeline poisoning by
automatically checking development phase isolation.
We envision a future where build system security checkers are as prevalent as
program code checkers.


\end{abstract}




\maketitle

\section{Introduction}%
\label{sec:introduction}


Open-source C software underpins society's critical software infrastructure,
including operating system kernels~\cite{linuxkernel,freebsd,xnu}, secure
communications tooling~\cite{openssh,openssl}, web and compute
servers~\cite{apachehttpd,lustre,openzfs}, and myriad other infrastructure
software~\cite{coreutils,xz-utils,systemd}.
Decades of work to secure the C programming language with bug
finding~\cite{chou01,reps98,schubert19}, program
verification~\cite{compcert,certikos,slam,boogie}, memory safe language
translation~\cite{3c,checkedc,rust,c2rust}, automated testing and
fuzzing~\cite{fioraldi_afl_2020,schumilo_kafl_2017,syzkaller}, and more make
implementation bugs harder for attackers to exploit.
But C software is not written only in C.
C codebases also include many additional languages for build automation, such as
autoconf~\cite{autotools}, Make~\cite{make}, and C preprocessor code~\cite{cpp}.
Such build code defines a C project’s structure and automates its configuration,
compilation, testing, packaging, and deployment~\cite{autotools,gnumake,cmake}.

A C project's build system is often a significant codebase with its own set of
security risks.
For instance, the Linux kernel's build system is over 200 thousand lines of
code~\cite{linuxkernel,gazzillo-esecfse21}.
Yet it is not only the size but the power of build automation that entails security
risk.
An attacker can exploit a C program without modifying a single line of C code,
 instead manipulating the build automation code that defines and controls how
the software is constructed, tested, and released.
Moreover, changes to build automation code can easily slip by developers, even
when software is checked with automated bug-finding, testing, or verification
tools, because build automation is not part of the program code per se.
For instance, the XZ Utils Backdoor (CVE-2024-3094~\cite{xzcve}) hid malicious
binary code in test cases and modified build code to inject the malicious
code into the libzlma compression library, resulting in a backdoor to publicly-released, popular distributions of
OpenSSH servers~\cite{xzcve}.
The attack illustrates \textit{poisoned pipeline execution}, one of “OWASP’s Top
Ten CI/CD\footnote{Continuous Integration/Continuous Development} Security Risks”~\cite{owasptoptencicd}, because it injects malicious
code via the build pipeline rather than directly in C program code.

\begin{figure*}
\begin{subfigure}{.45\linewidth}
    \centering
    \includegraphics[width=\linewidth]{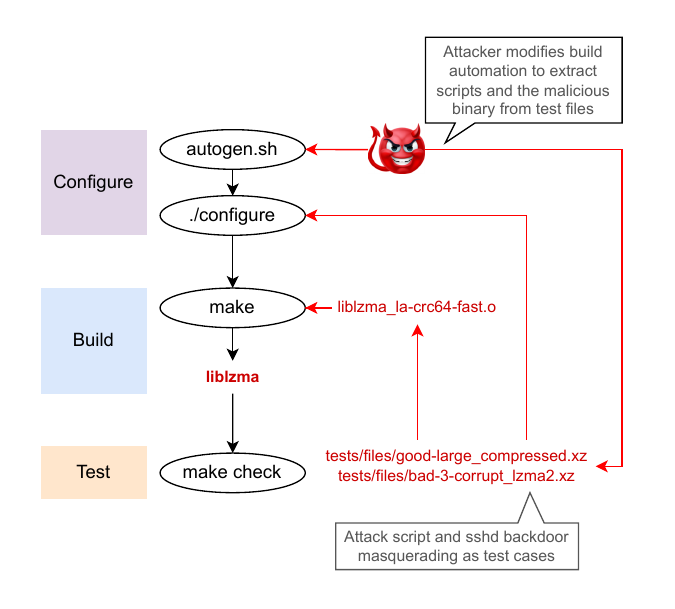}
    \caption{The XZ Utils backdoor injecting itself via pipeline poisoning.}%
    \label{fig:xz-utils-attack}
    \Description[
    Overview of how the XZ Utils backdoor injected itself into XZ Utils build
    pipeline.
    ]%
    {
    Overview of how the XZ Utils backdoor injected itself into XZ Utils build
    pipeline. The attacker first modifies the project's configuration phase to
    extract malicious code from poisoned test files. The configuration phase
    then modifies the project's build phase to build malicious code extracted
    from the poisoned test files into the liblzma library.
    }
\end{subfigure}
\begin{subfigure}{.45\linewidth}
    \centering
    \includegraphics[width=\linewidth]{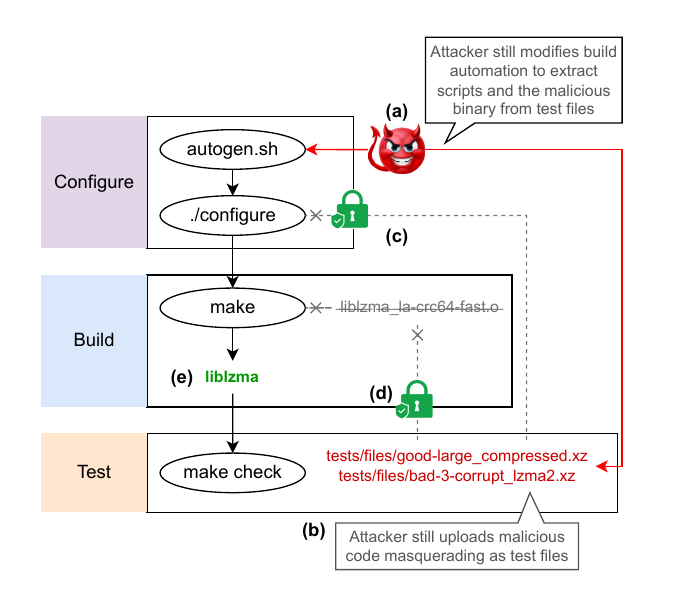}
    \caption{Phase isolation preventing the XZ Utils backdoor.}%
    \label{fig:xz-utils-prevented}
    \Description[
    Overview of how development phase isolation would prevent the XZ Utils backdoor.
    ]%
    {
    Overview of how development phase isolation would prevent the XZ Utils backdoor.
    As before, the attacker modifies the configure phase to extract malicious
    code from the poisoned test files. But now development phase isolation prevents
    the the configure phase from actually extracting code from the poisoned
    files. Moreover, development phase isolation also prevents the build phase from
    accessing the poisoned files, making it impossible for the build phase to
    compile the poisoned data into the liblzma library.
    }
\end{subfigure}
\end{figure*}

The problem is that developers do not secure build automation code as rigorously as program
code despite it being vulnerable to serious attacks.  
%
We posit this problem is partly because build code is not part of the software's program code per se, only having indirect control of the behavior of the actual software. 
%
%
%
%
%
%
%
%
But \emph{build code is still code}.
%
Therefore, we should be able to secure build systems by automatically analyzing build code just as we can secure programs by analyzing program code.
But analyzing build code comes with unique challenges compared with traditional program analysis.
Build systems are typically written in multiple, domain-specific languages designed to produce software~\cite{autotools,gnumake,felipec}.
Moreover, build systems, particularly those for C systems, employ metaprogramming extensively, generating several layers of build code to produce the final build system, making analysis more difficult.

Even if build code analysis tools were to exist, however, build systems lack analyzable specifications of their secure behaviors.
%
Developers use build systems to implement the software development
lifecycle as discrete phases~\cite{mcgraw,software-testing}, e.g., configure,
build, test, release, etc.~\cite{buildsystems,cisystems,cd}.
Many pipeline poisonings occur when data leaks into or out of a development
phase, or when one phase illicitly accesses information belonging to another
phase. 
For instance, the XZ Utils attack compiles code from the testing phase into the
final software~\cite{xz-utils}; the SolarWinds attack edits source code during
its build phase~\cite{solarwinds}, and Poisoned Pipeline Execution attacks
exfiltrate developer credentials during the testing phase~\cite{ppe}.
To defend against such attacks that illicitly leak or pass information among
development phases, we define a new build system security property, called
\emph{phase isolation}.
%
%
A build system satisfies phase isolation when each development phase executes under a well-defined set of access permissions, i.e., the principle of least privilege. 
A phase isolation checker ought to catch all the above poisoning attacks.

This paper outlines our research plan for designing analyses of build automation code and checkers for phase isolation.
As a proof of concept, we developed the first-of-its-kind development phase isolation checker, called \tool{}.
\tool{} collects file access patterns of the phases of a development pipeline, checking for isolation violations against a set of phase permissions.
It successfully identifies the insecure flow of malicious code from the test phase to the build phase on the compromised version of the XZ Utils codebase, demonstrating the feasibility of build code analysis to secure against pipeline poisoning.
We hope our research program will kickstart an era of powerful, automated protections for development pipelines and help secure critical software supply chains against increasing attacks on build automation code.

\section{Motivating example: the XZ Utils Backdoor}%
\label{sec:motivating-example}

The XZ Utils Backdoor~\cite{xzcve} was malicious code in the XZ Utils project
that created a backdoor to SSH daemons, specifically those patched to
use systemd~\cite{systemd}, which depends on the liblzma library built
by the XZ Utils repository.
The attackers hid backdoor code in plain sight in the public
codebase by masquerading it as a purported test
case~\cite{felipec,thesamesam,sentinelone}.
To obscure the linking of the backdoor code into liblzma, the attackers
poisoned the build automation pipeline by modifying the code that generates the
build system so that it that ultimately links the malicious code into the
liblzma library.

Figure~\ref{fig:xz-utils-attack} shows how the build automation was
poisoned.
The attackers uploaded the malicious backdoor, and scripts to inject it, as test
cases, since directly modifying C code would be obvious.
The attackers avoided modifying the Makefiles controlling the build
directly.
Instead, they exploited the common autoconf build code generation process,
which first generates a configuration script.
The configuration script inspects system properties and dependencies and then
itself generates the Makefile build system, which ultimately performs the
compilation.
The attackers modified an m4 macro in a file commonly included by autotools
called build-to-host.m4.
This macro passes commands to the shell to decompress and decrypt the first
test case, which contains a script to unpack another script, which ultimately
unpacks the malicious binary, called liblzma\_la-crc64-fast.o.
The Makefile, manipulated by the malicious scripts, then links the malicious
binary with the liblzma library.

The attack illustrates a lack of isolation between phases of the build.
A developer typically expects limited behavior of each phase of the build,
e.g., that the compile phase touches only source code and produces a binary,
while the test phase only reads test files and the resulting binary and
    produces a report.
In this case, attackers violated this expectation by using test files, not
source code or their object files, as part of the compilation phase and
modified the build system indirectly by altering the code that generates it.



Figure~\ref{fig:xz-utils-prevented} shows how building XZ Utils in a sandboxed
environment enforcing phase isolation prevents the backdoor
from being built into the liblzma library.
Just as before, the attacker modifies the project's configuration phase (a) to
inject malicious code from test case files (b) into the build phase.
Now, however, a development phase isolation check (c) prevents the configuration
phase from accessing the poisoned test files, since these files belong to the
test phase and not the configure phase.
Configuration then proceeds as normal, without the poisoned test files
manipulating the project's Makefile to inject their contents into the liblzma
library.
As a redundancy, another check (d) prevents the build phase from accessing
files from the test phase, making it impossible for the build phase to compile
and link the contents of the poisoned test files to the liblzma binary (e).
This example illustrates how phase isolation hardens build systems
against poisoning, even when malicious payloads are hidden in other parts of
the development pipeline.

\section{Research Stages}%
\label{sec:stages}

\begin{figure}
    \centering
    \includegraphics[width=\linewidth]{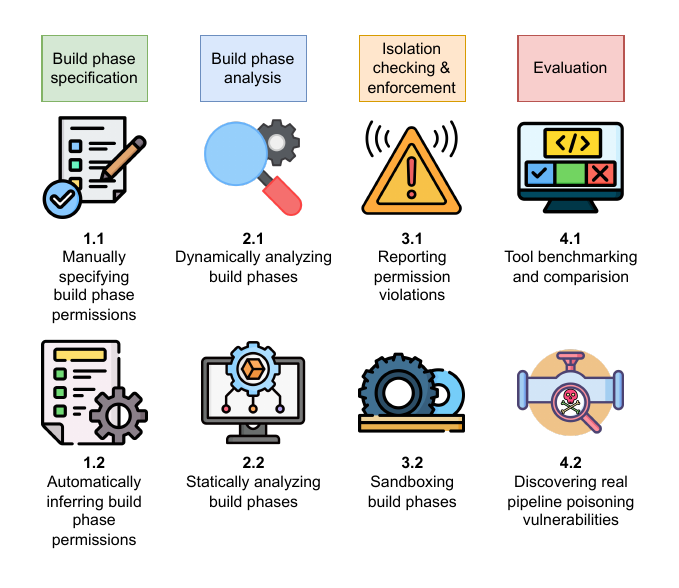}
    \caption{Development phase isolation research stages and tasks.}
    \label{fig:stages}
     \Description[
     Planned research stages and tasks for development phase isolation research.
     ]%
     {
     Our four stages of development phase isolation research.
     Each stage contains two research tasks.
     In the first stage we will specify development phase permissions.
     The tasks for stage one are manually specifying development phase permissions, and automatically inferring development phase permissions.
     In the second stage we will analyze development phase data and control flow.
     The tasks for stage one are dynamically analyzing development phase permissions, and statically analyzing them with phase simulation.
     In the third stage we will design tools to detect and prevent development phase violations.
     The tasks for stage three are developing tools to detect development phase violation permissions, and developing tools to enforce development phase permissions by running development phases in sandboxed virtual environments.
     In the fourth and final stage we will evaluate and use development phase isolation checking and prevention tools.
     In stage four task one, we will create a benchmark for comparing the effectiveness and performance of various tools for hardening development phases, and then use this benchmark to compare tools we develop in the previous research stages to other tools.
     In stage four task two, we will use the most effective tools, as determined by our benchmark results, to search for build system vulnerabilities in real-world software.
     }
\end{figure}

\newcommand{\task}[1]{Task~#1}

Figure~\ref{fig:stages} divides our research plan into
four stages. 
The first stage investigates the specification of development phase behavior.
The challenge is that, while there are conventions for software development life cycles, the precise set of phases and their expected behavior lack standardization.
Developers use a wide variety of automation tools and development models.
%
Therefore, securing development phases will first require specifying phase
commands and permissions.
We will design methods for manually and automatically defining phases permissions.
Unfortunately, it is not feasible for developers to manually specify development phase
permissions for projects with large and complex build systems, so in \task{1.2}
we will create tools to automatically infer phase specifications based on a
project's structure and pre-existing build system
standards~\cite{gnu-make-standard-targets}.

\begin{table}
\centering
\begin{tabular}{@{}lll@{}}
\toprule
Permissions                 & What is accessed      & Instrumentation                                                                                                      \\ \midrule
\multirow{2}{*}{Read/write} & Files                 & \begin{tabular}[c]{@{}l@{}}Run stat() system call \\ on files before \\ and after running\\ development phase\end{tabular} \\ \cmidrule(l){2-3}
                            & Environment variables & \begin{tabular}[c]{@{}l@{}}Run printenv -0 before\\ and after running\\ development phase\end{tabular}                     \\ \midrule
\multirow{2}{*}{Execute}    & Shell commands        & \begin{tabular}[c]{@{}l@{}}Check command \\ history after running \\ development phase\end{tabular}                        \\ \cmidrule(l){2-3}
                            & System calls          & \begin{tabular}[c]{@{}l@{}}Run strace() system \\ call before running \\ development phase\end{tabular}                    \\ \bottomrule 
\end{tabular}
\caption{Example access permissions and corresponding tracking instrumentation.}
\label{tab:permissions}
\end{table}

The second stage will investigate dynamic and static analyses of phase behavior.
In \task{2.1}, we plan to instrument build systems to detect permission
violations by tracking what information each phase accesses.
Table~\ref{tab:permissions} gives examples of data that development phase
instrumentation could track and potential techniques for tracking permissions.
Our proof-of-concept (Section~\ref{sec:preliminary-work}) shows that simply tracking file read/write accesses suffices to detect the XZ Utils poisoning attack.
Although dynamic analysis suffices for smaller projects, larger projects' development
phases can be time-consuming and computationally expensive to execute~\cite{llvm-compile-time}.
In \task{2.2} we will explore ways to make analyzing costly builds
more practical by statically predicting phase behavior with simulation.

In the third research stage we will develop tools using the algorithms in the
previous stage to check for and enforce development phase isolation.
In \task{3.1} we will create checkers to compare development phase permission
specifications to actual phase behavior, and report discovered permission
violations.
These warnings will enable developers to catch pipeline vulnerabilities and
poisonings before they reach production code.
Then in \task{3.2} we will go beyond detecting permission violations, and
execute development phases in sandboxed environments to prevent violations from
occurring entirely.
This is the method that Section~\ref{sec:motivating-example} uses to enforce
development permissions on the XZ Utils codebase and prevent the backdoor from being
injected into the project's binary outputs.

We will evaluate these techniques on real-world build systems.
%
First, in \task{4.1}, we will develop a benchmark of both safe and known vulnerable applications, drawing from both real-world examples of attacks~\cite{xz-utils,solarwinds,pytorch-attack,ledger-attack} and hand-crafted poisoning attacks in existing codebases.
%
%
We will use the benchmark evaluation in \task{4.2} to evaluate and improve our checkers as well as systematically check for the existence of build automation vulnerabilities in real-world code.
%
%
Our review will include code from a variety of sources, including GitHub's most
starred projects list, the GNU Free Software Directory, and projects considered
by prior research.
We will construct a standard build system review process to enable others to
easily replicate our findings, and report any vulnerabilities we find to
organizations such as the CVE Program to disseminate to the broader public.
We will also share any strategies we find for preventing discovered
vulnerabilities.

\section{Preliminary Work}%
\label{sec:preliminary-work}

\begin{figure}
    \centering
    \includegraphics[width=\linewidth]{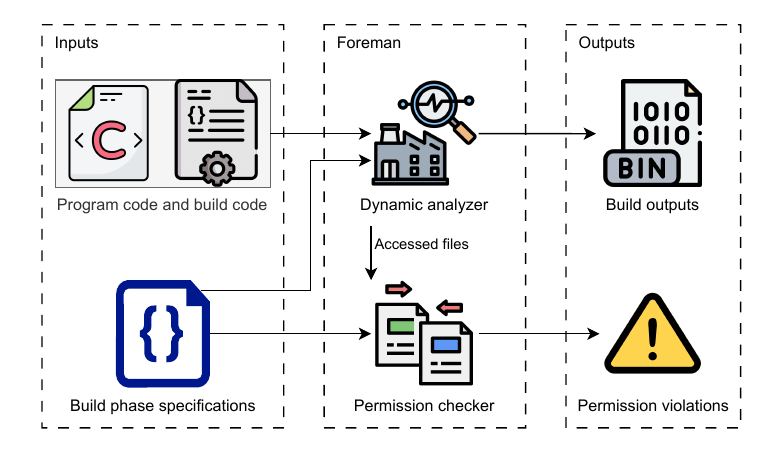}
    \caption{How \tool{} checks development phase isolation.}%
    \label{fig:foreman}
    \Description[
    \tool{} reads in a program and its permission specifications, dynamically
    detects when build phases access files that they are restricted from
    accessing, and reports these permission violations.
    ]%
    {
    \tool{} accepts two inputs: a program (including its source and build code),
    and a specification listing the commands and file permissions for each of
    the program's build phases. \tool{} runs the build phases while keeping
    track of which files each phase accesses. \tool{} then checks whether each
    phase accesses a file that the specification restricts it from accessing.
    Each time a phase accesses a restricted file, \tool{} reports the access as
    a permission violation.
    }
\end{figure}

To show how development phase isolation can effectively prevent pipeline poisoning,
we implemented a proof-of-concept checker called \tool{}.
%
%
\tool{} wraps development phases and monitors their file accesses.
Even with a simple implementation comprising only 177 lines of Python code, \tool{} is already capable of detecting the pipeline poisoning form the XZ Utils backdoor. 

%


Figure~\ref{fig:foreman} is the architecture of \tool{}. 
%
It takes a specification of the commands to run development phases and their expected permissions as input.
Currently, we create this specification manually as a JSON file.
Given the commonalities between build automation code~\cite{gnu-make-standard-targets}, we hope to reduce the burden of specifying this via automated build specification analyses as described in Section~\ref{sec:stages}.

Next, \tool{} dynamically analyzes the build system by running each build phase
and recording which files they access.
\tool{} determines if a build phase accesses a file by running the Linux
\texttt{stat()} system call both before and the phase runs; if the post-phase
access time is more recent, then \tool{} judges the phase as having accessed
the file.
%
%
Finally, \tool{} checks the actual file access patterns against the development phase specification. 
If a build phase accesses a file that the project's specification restricts
that phase from accessing, then \tool{} reports the file access as a permission
violation by logging its details to an output file.

\subsection{Case Study: Detecting the XZ Utils Backdoor}
\label{subsec:case-study}

\begin{lstlisting}[
    style=json-txt,
    label={lst:xz-spec},
    caption={JSON specification of XZ Utils build phases and file permissions.},
    escapechar={^}
]
[
 {"name":"autogen",  "commands":[["./autogen.sh"]]},
 {"name":"configure","commands":[["./configure"]]},
 ^\highlightcode{\{  "name": "compile", "commands": [["make"]],}^
 ^\highlightcode{"restricted": ["tests/files/"] \}}^
]
\end{lstlisting}


Listing~\ref{lst:xz-spec} is the minimal specification we defined for \tool{}
to detect the poisoned files.
%
The highlighted portion on Lines~4-5 defines XZ Util's \texttt{compile}
phase, which runs \texttt{make} to compile the project and is restricted from
accessing files in the directory \texttt{tests/files/}, a reasonable restriction as source and test files are typically separate.
%
%



\noindent
\begin{minipage}{\linewidth}
\begin{lstlisting}[
    style=json-txt,
    label={lst:xz-warnings},
    caption={\tool{} warnings about poisoned XZ Utils test files, with file
    paths shortened to file names.}
]
warning: compile accessed bad-3-corrupt_lzma2.xz
warning: compile accessed good-large_compressed.lzma
\end{lstlisting}
\end{minipage} 

Listing~\ref{lst:xz-warnings} shows the warnings \tool{} emits after building
the compromised version of XZ Utils with the build phases specified 
in Listing~\ref{lst:xz-spec}.
\tool{} warns that the \texttt{compile} phase violates its specified
permissions by accessing the files \texttt{bad-3-corrupt\_lzma2.xz} and
\texttt{good-large\_compressed.lzma}; which are the exact files that contain the malicious code used in the attack~\cite{russ-cox-xz-utils}.
%
%
This approach shows that no specific defenses against the XZ Utils backdoor mechanism itself are needed; just by enforcing development phase isolation, \tool{} detects the violation that XZ Utils backdoor depends on for the attack.
%


%
%


\vspace{1em}

\section{Future Plans}%
\label{sec:future-plans}

%





\tool{} illustrates the feasibility of phase isolation checking and leads to several concrete next steps.
%
%
Firstly, \tool{} depends on a manual specification of phase permissions.
%
The problem with manual specification is that if a developer incorrectly specifies a build phase's permissions, \tool{} may miss vulnerabilities. 
%
We aim to mitigate this issue by taking advantage of commonalities between
build phases (e.g., separate directories for source code and test code) to
automatically infer build phase permissions.
Additionally, an overly-strict specification can trigger false alarms. 
For example, in this case study we only restricted XZ Utils' \texttt{compile}
phase from accessing test files, but if we had also restricted its
\texttt{configure} phase from accessing test files, then we would have
received dozens of false alarms, as the \texttt{configure} phase does in fact
access the project's test files.
Code analyses generally have trade-offs between precision and recall~\cite{precision-vs-recall}, and
%
we plan to ameliorate these issues by exploring both static and dynamic build
phase isolation checking analyses, so that developers can choose an approach that best meets their needs.



We plan to continue our work on development phase isolation checking by
following the stages outlined in Section~\ref{sec:stages}, beginning with the
automatic inference of build system specifications.
%
\tool{}'s file access tracking is coarse in that it only allows users to
specify which files build phases may and may not access in any way.
However, some build phases may only need permission to perform specific actions
on a file like reading or writing; for instance, a compilation phase may need
permission to read source files but not to write to them.
We intend to extend \tool{} in the future with mechanisms allowing users to
specify such higher-resolution permissions, both statically and dynamically.
Additionally, to make \tool{} a complete development phase isolation checker, we
aim to augment it with static and dynamic checks on build phase command
execution permissions.
%
%
%
Finally, we plan to make phase isolation checkers easier to use and integrate them
into existing build pipelines, so that we can perform large scale evaluation of real-world build pipelines to find new vulnerabilities. 
While we initially focus on C software's build system, we expect the phase isolation checkers to be applicable to other build system languages by porting the analysis algorithms.
Ultimately, we hope for a future where build code checkers are as ubiquitous and powerful as program
code checkers.


\begin{acks}
We would like to thank to anonymous reviewers for their feedback.  This work was supported in part by NSF grant CCF-1941816.
\end{acks}

\balance{}
\bibliographystyle{plain}
\bibliography{references.bib}

\end{document}